\documentstyle[newarcrc,fleqn,epsf]{article}

\newcommand{\mk}{$M_{\rm K}$}
\newcommand{\sk}{$S_{\rm K}$}
\newcommand{\vi}{$V-I_c$}
\newcommand{\vk}{$V-K$}
\newcommand{\Ftio}{$T_{\rm TiO}$}
\newcommand{\Msun}{$M_{\odot}$}

\newcommand{\tio}{TiO$\lambda7500,7165$}
\newcommand{\od}{$\bigcirc \hspace*{-1.68ex}\bullet$\,}
\newcommand{\odia}{$\Diamond$\hspace*{-1.44ex}\raisebox{0.2ex}{$\bullet$}\,}

\title{Secondary stars in CVs -- the observational picture}

\author{K. Beuermann \address{Universit\"ats-Sternwarte
G\"ottingen\\Geismarlandstr. 11, 37083 G\"ottingen, Germany\\}
\thanks{Research supported in part by the Bundesministerium
f\"ur Bildung und Wissenschaft under BMBF/DLR grant
No. 50\,OR\,9210.}}

\begin{document}

\maketitle

\begin{abstract}
Recent theoretical and observational progress has substantially
improved the definition of the lower main sequence and established a
new basis for a comparison of main sequence stars and the secondaries
in CVs. The evolutionary sequences of Kolb \& Baraffe [1999] imply
that the secondaries in many CVs are expanded compared with main
sequence stars of the same mass as a consequence of unusually high
mass transfer rates and/or pre-CV nuclear evolution.  We show that the
location of the secondaries of all well-studied CVs in the spectral
type period diagram implies that they are consistent with having
near-solar metallicities. We show, furthermore, that the surface
brightness of K/M stars depends on gravity and metallicity and present
new Barnes-Evans relations valid for dwarfs of near-solar metallicity
and the secondaries in CVs of the galactic disk population. Distances
derived by the surface brightness method agree with recent
measurements of the trigonometric parallaxes of a few selected
systems.
\end{abstract}

\section{Are the secondary stars in CVs main sequence stars?}

The secondary stars in CVs have been recognised to follow
approximately the mass--radius relation of main sequence field stars
[Echeverr\'{\i}a, 1983, Patterson, 1984, Warner 1995, Smith \&
Dhillon, 1998]. A detailed comparison is complicated by the lack of
information both on the masses of the secondary stars in CVs and of
the masses and radii of main sequence field stars. It is advantageous,
therefore, to restrict the comparison to quantities readily observable
in CVs, as the orbital period $P$ and the spectral type $Sp$ of the
secondary star [Beuermann et al., 1998, henceforth BBKW98, Smith \&
Dhillon, 1998]. This approach requires to construct an equivalent
diagram for field stars predicting the orbital period of a CV at which
a field star of given spectral type would fill the Roche-lobe of the
secondary. Note that this prediction is almost independent of the mass
of the white dwarf primary.

\begin{figure}[htb]
\begin{minipage}[t]{78mm}
\mbox{\epsfxsize=77mm\epsfbox{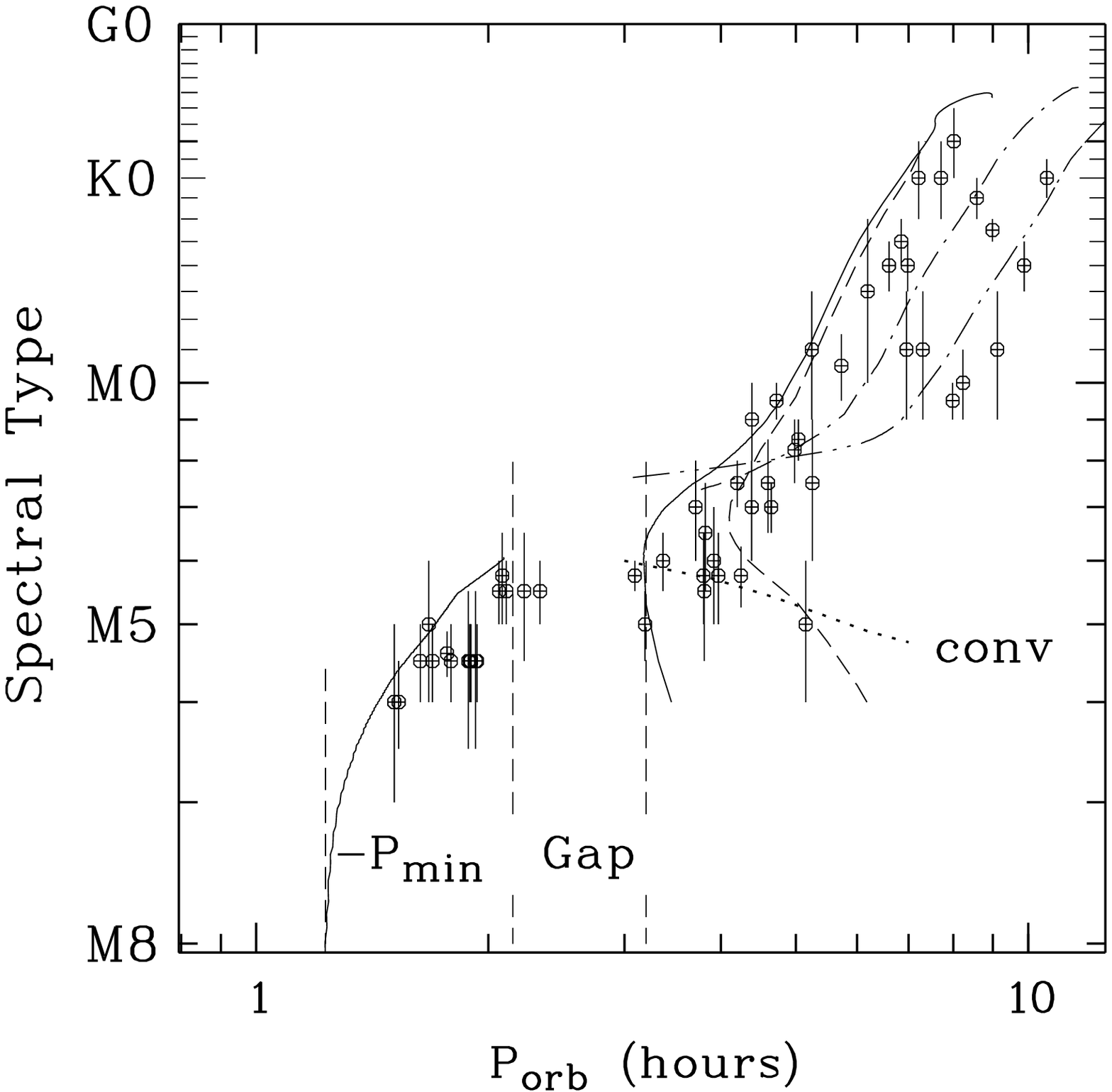}}
\vspace*{-3mm}\caption{Spectral types of CVs as a function of orbital
period. The theoretical curves are from BBKW98 and Kolb \& Baraffe
[1999].}
\end{minipage}
\hspace{\fill}
\begin{minipage}[t]{78mm}
\mbox{\epsfxsize=77mm\epsfbox{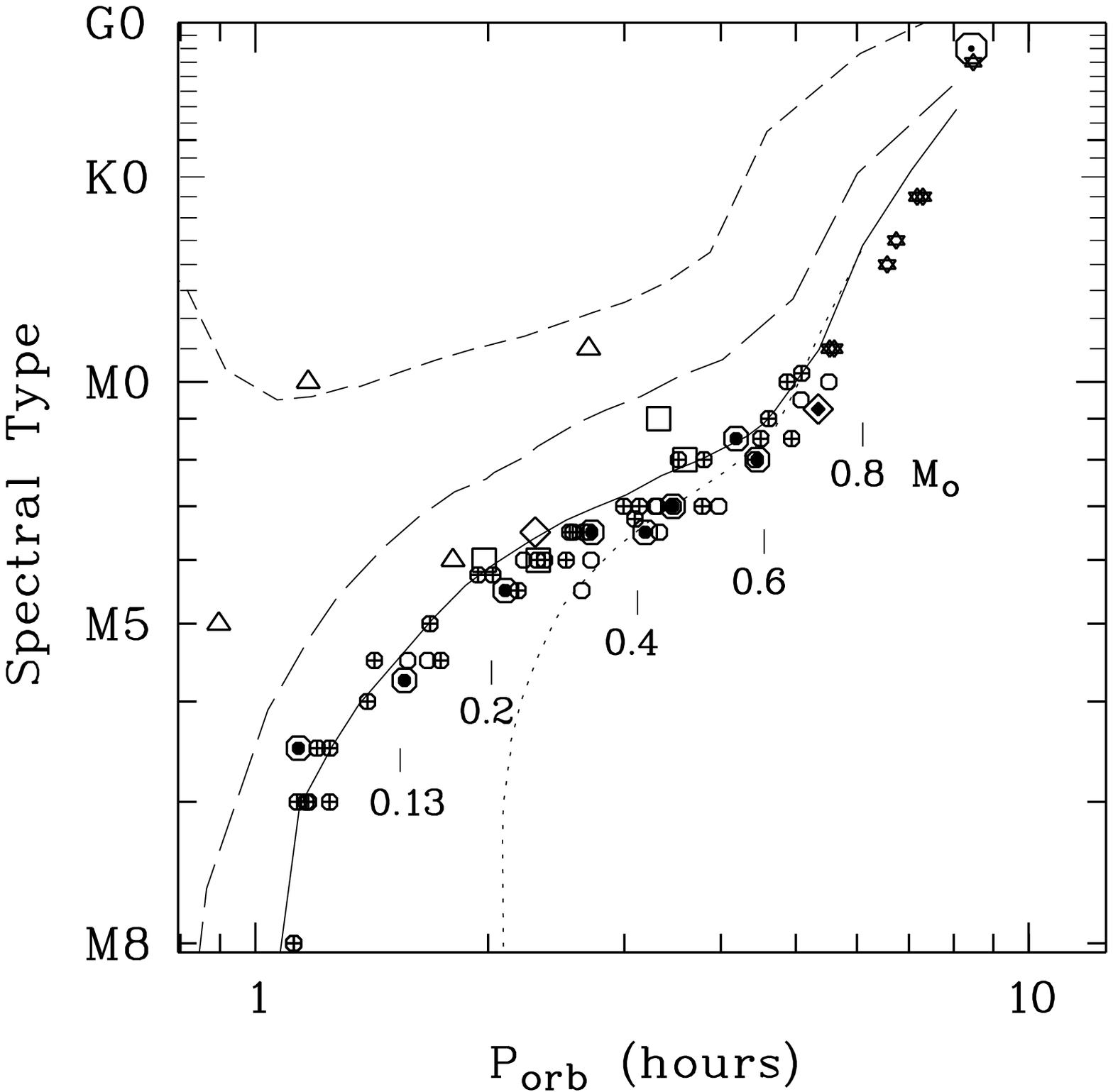}}
\vspace*{-3mm}\caption{Spectral type orbital period distribution for
field stars assumed to be Roche-lobe filling secondaries in CVs. See
text for explanation of symbols and theoretical curves [BCAH98].}
\end{minipage}
\end{figure}

Figure 1 shows the spectral types of the secondaries in CVs with known
orbital period. The data points are based on published optical/near-IR
spectra quoted in the compilation of Ritter \& Kolb [1998] with some
recent spectral type determinations added [e.g., Schwarz et al., 1999,
Smith et al., 1999, Thomas et al., 1999, Thorstensen et al.,
1999]. Purely photometric spectral type assignments have not been
included. We have also not included spectral types derived from IR
spectra only.

Constructing a spectral type period diagram for field stars requires
knowledge of their masses and radii. Unfortunately, both quantities
are not generally well known. In this situation it is important to
note that recent progress in the construction of stellar models
[Baraffe et al., 1998, henceforth BCAH98] combined with the {\it
NextGen} model atmospheres of Hauschildt et al. [1999] have led to a
substantially improved definition of the lower main sequence [Leggett
et al., 1996, henceforth L96]. In fact, as noted by BBKW98, the
theoretical and observational radii of M-stars [L96] agree at the few
\% level for stars of given absolute $K$-band magnitude \mk{}.  Using
the observed radii, we have established a radius scale which allows to
estimate the radii of field stars of given \mk{} [Beuermann et al.,
1999, henceforth BBH99]. For the present purpose, we adopt the same
stellar models which accurately reproduce the radii to estimate the
masses, too. Specifically, we use the theoretical relation between
mass and absolute $K$-band magnitude, $M(M_{\rm K})$ [BCAH98] to
calculate the above defined `orbital period' for a field star of known
\mk{} and $Sp$. We do not follow the approach of Clemens et al. [1998]
who adopted the observational relation between mass and absolute
magnitude of Henry \& McCarthy [1993] because this relation represents
a mean for stars of different age and metallicity and we want to study
how differences in metallicity affect the location of stars in the
$Sp-P$ diagram.

Figure 2 shows the resulting $Sp-P$ diagram for field stars of
spectral types K/M supplemented by the Sun. The main sequence of stars
with near-solar metallicity is delineated by eight young disk (YD)
field stars from L96 (\od), the YD binary YY Gem (\odia), the Sun
($\odot$), and further 56 stars ($\star, \circ,\oplus$) with radii
from [BBH99]. The effects of decreasing metallicity are indicated by
the binary CM Dra ($\Diamond$) and by four old disk stars ($\Box$) and
four halo stars ($\bigtriangleup$) from L96. Also shown in Fig. 2 are
the models curves from the BCAH98 stellar models, namely the ZAMS
model for solar metallicity (solid line), and models for stars aged
$10^{10}$ yrs with 1/3 solar metallicity (long dashes) and 1/30 solar
metallicity (short dashes). For the model stars, the spectral type is
assigned by converting the theoretical colour $I-K$ to $Sp$
[BBKW98]. The solid curve is expected to agree with the locus of the
late-type YD stars.  The slight displacement is probably due to
remaining errors in the transformations used.

Comparison with Fig. 1 indicates that the (zero age) main sequence of
low-mass field stars with near-solar metallicity coincides with the
locus of CV secondaries of the earliest spectral types at any given
value of $P$. The upper left in Fig. 1 is devoid of CVs and shows that
none of the systems included in the figure contains a secondary with
metallicity substantially lower than the Sun (larger than solar
metallicities are not excluded but remain unproved at present). The
low space density of Pop\,II CVs is in agreement with population
studies by Stehle et al. [1999].  
The secondaries in many CVs with $P > 3$\,h have a later spectral type
than expected for ZAMS secondary stars of near-solar metallicity.
These secondaries are expanded over and have lower masses than
Roche-lobe filling ZAMS stars. The two principal causes for this
expansion are the loss of thermal equilibrium due to the on-going mass
transfer and nuclear evolution prior to the onset of Roche-lobe
overflow. Kolb \& Baraffe [1999] have computed corresponding
evolutionary sequences which nicely explain the observed behaviour and
of which first results were presented in BBKW98. The paths included in
Fig. 2 refer to secondaries starting as ZAMS stars of 1\,\Msun{} and
evolving under mass loss rates of $1.5\times10^{-9}$\,\Msun\,yr$^{-1}$
(solid curve) and $5\times10^{-9}$\,\Msun\,yr$^{-1}$ (dashed
curve). The point on these paths where the secondary becomes fully
convective is indicated by the dotted curve [Kolb \& Baraffe,
1999]. Late spectral types in CVs with orbital periods between 3 and 5
h can be explained by this scenario.  The evolutionary models suggest
that whenever the secondary becomes convective it has reached
$\sim0.2$\,\Msun{} and $Sp \sim $M4.5. If angular momentum loss drops
rapidly at this point the secondary enters the period gap and
re-appears below the gap with the same mass and spectral type $Sp \sim
$M4.5.
The comparatively late spectral types in CVs with orbital periods
larger than 5 h can not be explained by the loss of thermal
equilibrium due to mass loss but are consistent with nuclear evolution
of the secondary star prior to the onset of mass transfer. Two
evolutionary paths are included in Fig. 2 for secondaries starting
mass transfer at $M = 1$\,\Msun{} with a rate of
$1.5\times10^{-9}$\,\Msun\,yr$^{-1}$ and a central hydrogen fraction
which is reduced to 0.16 (dot-dashed curve) or is practically
exhausted (dot-dot-dashed curve) [BBKW98, Kolb \& Baraffe,
1999]. Early evolutionary calculations of a similar type were
performed by Whyte \& Eggleton [1980].

The model calculations of Kolb \& Baraffe [1999] indicate that stars
driven out of thermal equilibrium stay at approximately the same
effective temperature and spectral type as an undisturbed star of the
same mass. Their results allow to estimate the masses of the secondary
stars and thereby their Roche radii from the observed spectral
types. This is a prerequisite for distance determinations of CVs using
the surface brightness method [Bailey 1981].

\section{The brightness distribution of CV secondaries}

The Roche-lobe filling secondary stars experience gravity darkening
and will not be of uniform surface brightness. More seriously and more
difficult to model are the variations in surface brightness caused by
irradiation. Marsh [1990] showed that the strengths of the TiO bands
and the NaI$\lambda 8183,8195$ absorption-line doublet decrease on the
hemisphere of the M5 secondary star in HT Cas which faces the
primary. Naively, one might expect that heating of the secondary star
produces an earlier spectral type and a corresponding increase in the
TiO band strength. Observation indicates an opposite result, a
decreased flux of the TiO features on the illuminated hemisphere,
which suggests that a major change in the atmospheric structure takes
place as a result of heating. Schwope et al. [1999] present a nice
tomographic picture of the non-uniform appearance of the secondary
star of QQ Vul in the NaI lines.  Similar results have been obtained
e.g. for AM Her [Southwell et al., 1995], illustrating the
complications caused by the non-uniform appearance of the secondary
star for dynamic mass determinations. This non-uniformity must be
taken into account when applying the surface brightness method for
distance measurements.

\begin{figure}[t]
\begin{minipage}[t]{80mm}
\mbox{\epsfxsize=8.0cm\epsfbox{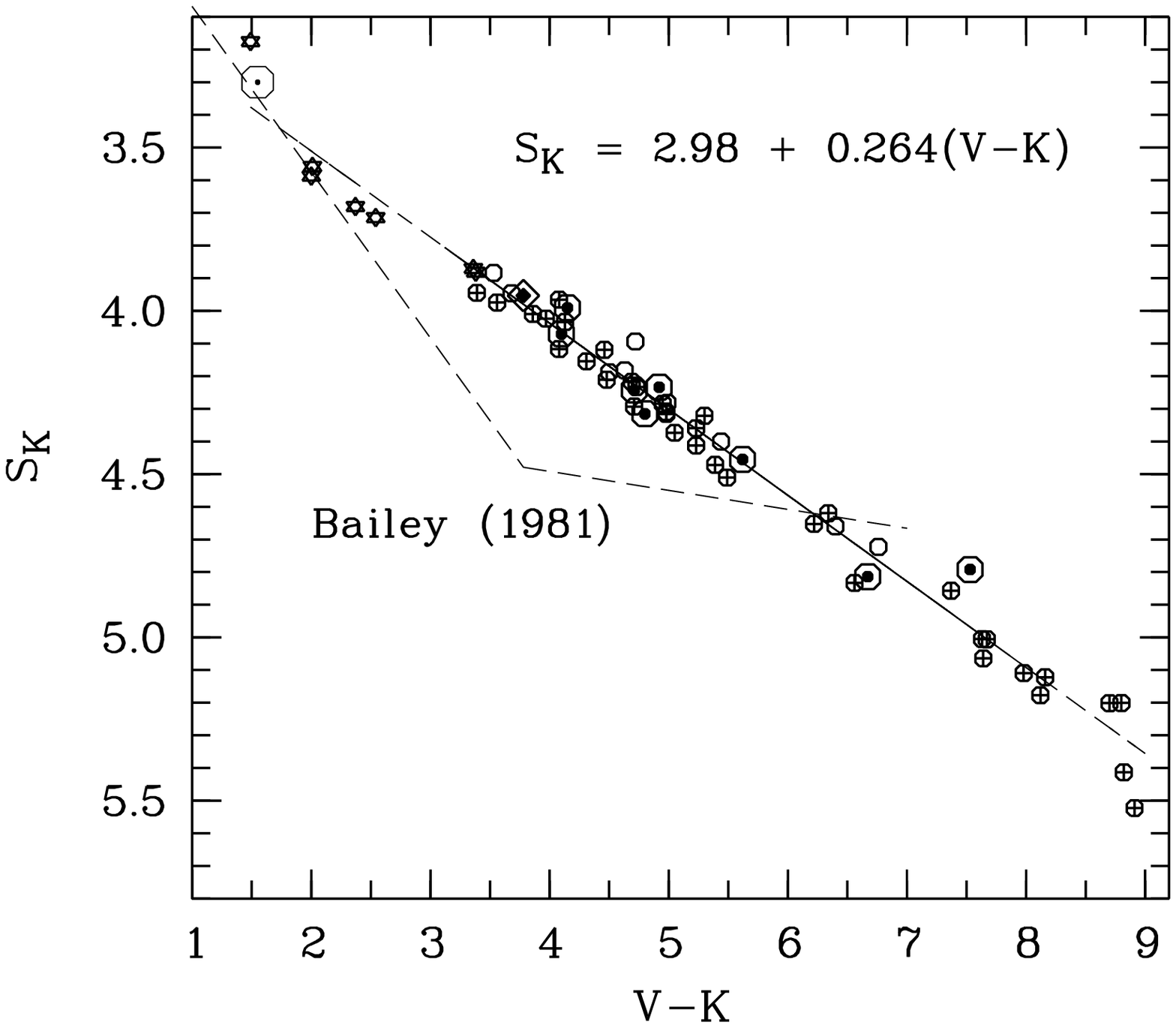}}
\vspace{-1.0cm}\caption{Surface brightness \sk{} in the $K$-band as a
function of colour \vk{} for single field stars with near-solar
metallicity.}
\end{minipage}
\hspace{\fill}
\begin{minipage}[t]{76mm}
\mbox{\epsfxsize=7.6cm\epsfbox{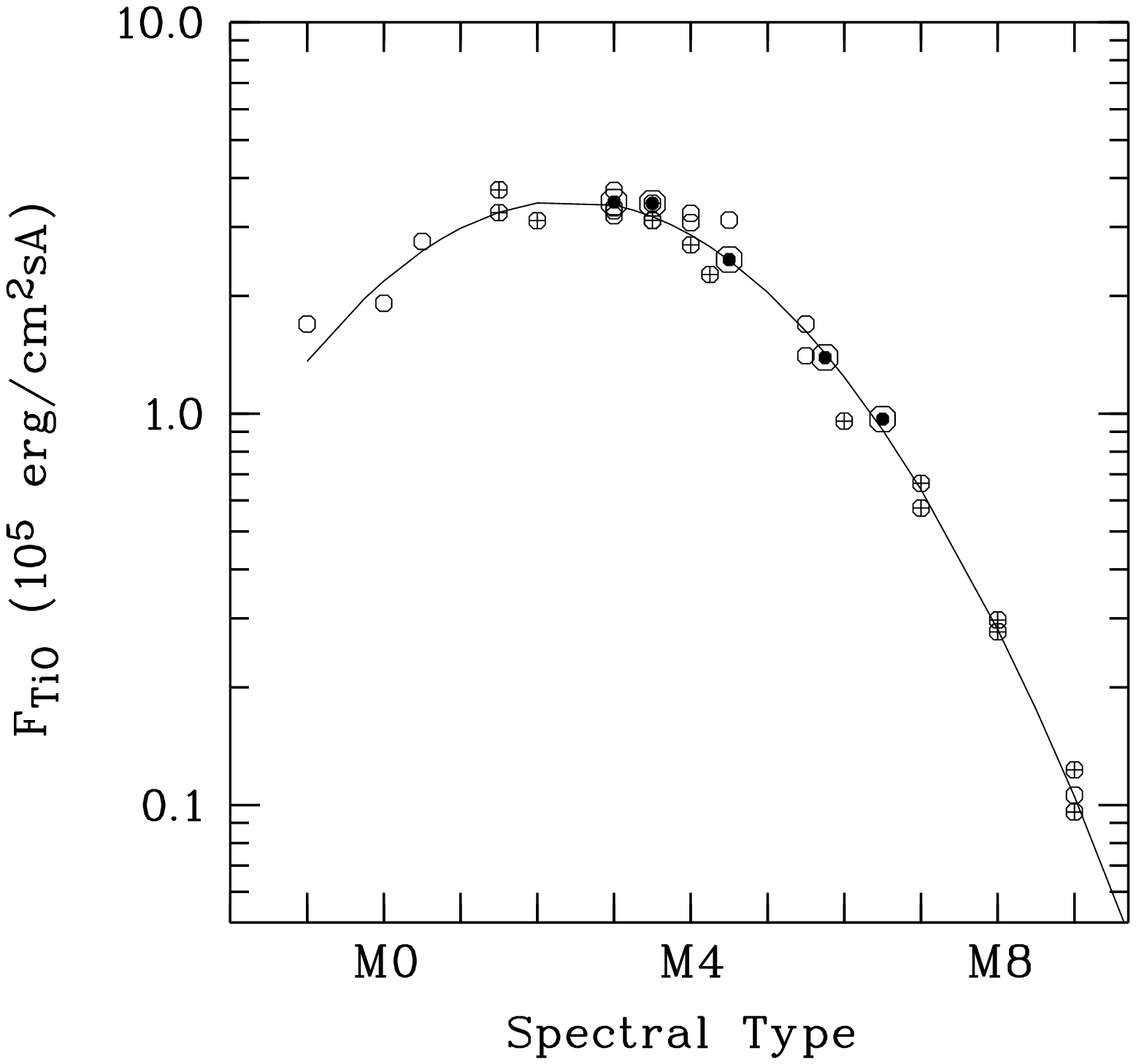}}
\vspace{-1.0cm}\caption{Surface brightness \Ftio{} (in $10^5$
erg\,cm$^{-2}$s$^{-1}$) as a function of spectral type $Sp$.}
\end{minipage}
\end{figure}

\section{Barnes-Evans relation for field M-dwarfs}

Barnes \& Evans [1976] found that the visual surface brightness of
giants and supergiants (the only stars for which directly measured
radii are available, apart from the few eclipsing binaries) is a
function of colour only and independent of luminosity (gravity).
BBH99 derived the surface brightness for the M-dwarfs studied by L96
and demonstrated that it deviates from that of giants. The implied
small gravity dependence is in perfect agreement with recent model
calculations for dwarfs and giants [BCAH98, Hauschildt et al.,
1999]. The dwarf/giant difference reaches a maximum for $Sp \simeq
\,$M0 (\vi\,=\,1.7, \vk\,=\,3.5). BBH99 also showed that the surface
brightness of dwarfs in the $K$-band depends on metallicity in
agreement with the predictions of BCAH98. Since the well-observed CVs
have near-solar metallicities (Figs. 1 and 2), consideration of the
gravity and metallicity dependencies allows improved surface
brightness-colour relations to be established which are valid, e.g.,
for CVs of the galactic disk population. Fig. 3 shows the resulting
surface brightness \sk{} in the $K$-band as a function of colour \vk{}
which can be fit by the linear relationship given in the figure. This
relation differs from the original relation of Bailey [1981] which was
widely used in CV research. Bailey's calibration of \sk{} depended on
the Barnes-Evans relation for giants and is, therefore, expected to be
low by some 0.3-0.4 mag near spectral type M0 or \vk$ = 3.5$. The
systematic difference between the two relations is primarily due to
the gravity dependence of \sk. The often-cited near constancy of \sk{}
for M-dwarfs does not exist.  The scatter in our data is substantially
reduced over that in Bailey's diagram, and in the similar one by
Ramseyer [1994], because we considered only single dwarfs with
near-solar metallicity and avoided colour transformations between
different photometric systems. Low-metallicity M-dwarfs have \sk{}
values lower by $\sim0.5$ mag. There is a systematic uncertainty in
the derived surface brightness of about 0.1 mag carried over from the
remaining uncertainty in the radius scale [BBH99].

The secure identification of CV secondary stars requires spectroscopic
observations in the optical/near-IR spectral regions. M-stars are best
detected by their pronounced TiO bands in the red part of the optical
spectrum which display a variation in band strength ratios with
spectral type [Wade \& Horne, 1988, Marsh, 1990]. We have calibrated
the {\it absolute strength} of the flux depression at 7165\,\AA{}
relative to the quasi-continuum at 7500\,\AA{} vs. spectral type for a
selection of field stars with near-solar metallicities and show the
result in Fig. 4.  The quantity $F_{\rm TiO} = (F_{\lambda7500} -
F_{\lambda7165})\times d^2/R^2$ is the TiO flux depression expressed
as a surface brightness, with $d$ the distance and $R$ the stellar
radius. The spectral fluxes $F_{\lambda7500}$ and $F_{\lambda7165}$
represent averages over $\pm 50$ and $\pm 25$\AA,
respectively. Different from the surface brightness \sk{} above,
$F_{\rm TiO}$ is given in physical units. A second order polynomial
fit to the logarithm of $F_{\rm TiO}$ is 
\begin{equation}
F_{\rm TiO} = 10^{5 + \alpha}~{\rm erg\,cm}^{-2}{\rm s}^{-1}{\rm \AA}^{-1}
\qquad {\rm with} \qquad \alpha = -1.477 + 0.5332\,(10-S) -
0.03516\,(10-S)^2
\end{equation}
where $S$ is the spectral subtype for M-dwarfs, i.e. $S = 5.5$ for an
dM5.5 secondary, and $S = -1$ for K7, preceding M0. Eq. (1) is
applicable to all CVs which have secondaries with near-solar
abundances. For metal-poor M-dwarfs of the old-disk and halo
populations, the surface brightness $F_{\rm TiO}$ is reduced over that
of Eq. (1).  In extreme subdwarfs the $\lambda7165$ flux depression
disappears and our definition of $F_{\rm TiO}$ becomes
meaningless. Hence, Eq. (1) needs revision should spectral types
become available for the secondaries in Pop II CVs, e.g., in globular
clusters.

\begin{center}
\begin{table}[t] 
\caption[]{Distances of CVs determined from the strength of the \tio{}
band strength, the $K$-magnitude of the secondary and compared with
trigonometric parallaxes. The distances scale with secondary mass $M_2$ as
$M_2^{1/3}$. }
\begin{tabular*}{\hsize}{@{\extracolsep{\fill}}lclcccccc}
\noalign{\smallskip} \hline \noalign{\smallskip} 
Name & $P_{\rm h}$ & $Sp$ & $f_{\rm TiO}$ & $K$ & $M_2/M_{\odot}$ &
$d_{\rm pc}(f_{\rm TiO})$ & $d_{\rm pc}(K)$ & $1/\pi$\\
\noalign{\smallskip}\hline\noalign{\smallskip} 
\multicolumn{9}{l}{\it (1) \qquad The clear cases :}\\[0.2ex]
V834Cen & 1.69 & M5.5 & 1.78 & $13.85$ &0.13&$113\pm16$ & $111\pm9$ & \\
Z Cha & 1.79 & M5.5 & 1.27 & $14.05$ & 0.13&$128\pm18$ & $117\pm10$ & \\
AM Her & 3.09 & M4+ & 21.0& $11.71$ & 0.27 &$81\pm7$&$90\pm7$&$85\pm5$\\
U Gem & 4.25 & M4+ & 37.0 &$10.91$ &0.41&$88\pm11$ & $92\pm7$ & $96\pm4$\\
[0.5ex]
\multicolumn{9}{l}{\it (2) \qquad Some hopefully clear cases :}\\[0.2ex]
BL Hyi & 1.89 & M5.5 & 1.13 &---& 0.16 & $163\pm23$ & &\\
RX0203 & 1.89 & M2.5 & 1.10 &---& 0.45 & $630\pm60$ & &\\
UZ For & 2.11 & M4.5 & 0.89 &---& 0.20 & $263\pm29$ & &\\
TT Ari & 3.30 & M3.5 & 2.20 &---& 0.30 & $297\pm25$ & &\\
IX Vel & 4.66 & M2?  & ---  &10.7&0.50 &  &$103\pm12$  & $96\pm9$\\ 
AE Aqr & 9.88 & K4 &---& $>8.73$ & 0.50 & & $>87\pm10$ & $102\pm32$ \\[0.5ex]
\multicolumn{9}{l}{\it (3) \qquad Some controversial cases :}\\[0.2ex]
HT Cas & 1.77 & M5.4 & 0.98 & $15.4:$ & 0.13& $150\pm15$ &$>215$&\\
SS Cyg & 6.60 & K4 &---& $10.2:$ & 0.70 &---& $152\pm15:$ & $166\pm13$\\
V1309 Ori & 7.98 & M0.5 & 1.09&$>15.5$& 0.45 & $745\pm70$ & $>1370$&\\[0.5ex]
\noalign{\smallskip}\hline\noalign{\smallskip}
\end{tabular*}
\vspace{-10mm}
\end{table}
\end{center}

\section{Distance measurements of CVs}

The surface brightness method [Bailey, 1981] allows the measurement of
distances of CVs if the $K$-band flux of the secondary star or its
spectral flux in the \tio{} band can be measured. To be sure, the
secondaries in CVs display a non-uniform surface brightness and care
should be taken if only a single spectrum or flux measurement are
available. Ideally, the full orbital modulation is needed in order to
select the most appropriate view on the secondary star for comparison
with the appropriate surface brightness of field stars. In general,
the $K$-band fluxes are not seriously affected by irradiation, while
spectral features as the TiO bands and the NaI absorption lines are
greatly diminished in strength on the heated face of the secondary
star [e.g. Marsh, 1990, Schwope et al., 1999]. Both methods have,
therefore, advantages and disadvantages. The advantage of using the
TiO band strength lies in its easy identification in observed spectra.

We have used both the $K$-band brightness of the secondary and its TiO
band strength to derive the distances of selected CVs and show some
results in Table 1 (for a more detailed discussion see Beuermann \&
Weichhold, 1999. Note that the error in $M_2$ has not been considered
and the distances scale as $M_2^{1/3}$). Ideally, both methods should
yield the same distances which should agree within errors with the
trigonometric parallax. This is, in fact, the case for of AM Her and U
Gem (parallaxes by C. Dahn, private communication, Harrison et al.,
1999). We have confidence also in the distances derived for V834 Cen
and Z Cha which have no trigonometric parallaxes yet. There are some
hopefully clear cases which represent CVs with well-defined TiO
features. Obtaining independent distance information for these
systems, e.g., from the $K$-band fluxes of the secondary stars is
clearly desirable. There are also controversial cases, however, which
indicate some of the pitfalls of the methods.  For HT Cas, the
distance derived from infrared photometry (Berriman et al., 1987) is
probably wrong, as noted already by Ramseyer [1994]. Another important
case is that of SS Cyg which has $K = 9.4$ in quiescence. If that IR
flux is entirely due to the K4 secondary, as assumed by Bailey [1981],
our calibration yields $d = 105$\,pc. On the other hand, the secondary
contributes only about $\sim 50$\% of the visual flux which implies
that it is has $K\,\simeq \,10.2$ and raises the distance to
$d\,\simeq \,152$\,pc, consistent with the recent HST FGS parallax of
166 pc [Harrison et al., 1999]. Note that assuming a K5 secondary
would reduce $d$ again to 133\,pc. Finally, the 8-h AM Herculis binary
V1309 Ori has an (evolved) M0$-$M1 secondary and a TiO distance of 745
pc. The optical flux and spectral type imply that the secondary
accounts for most of the $K$-band flux which is inconsistent with the
statement of Harrop-Allin et al. [1997] of a contribution as low as
$\sim 20$\% based on $K$-band spectroscopy of the illuminated face of
the secondary. The discrepancy is likely due to problems associated
with the interpretation of illuminated stellar atmospheres.

We conclude that reliable distances of CVs can be obtained from both
the $K$-magnitude and the TiO band strength of the secondary if its
flux contribution can be unequivocally determined, illumination
effects are taken properly into account, and the viewing direction on
to the secondary is known. Measuring accurate distances to CVs allows
to derive basic quantities as their absolute magnitudes and the mass
transfer rates [Warner, 1987]. Much of our understanding of the
physics of CVs [Warner, 1995] is based on the tedious derivation of
such quantities.

\section{Acknowledgements} 
I thank Isabelle Baraffe and Ulrich Kolb for many discussions and for
allowing me to show their evolutionary paths in Fig. 1 prior to
publication, Hans Ritter for pointing out some errors in my earlier
compilation of spectral types of secondary stars, and Marc Weichhold
for the initial work on this subject.

\end{document}